# An agent-based evaluation of impacts of transport developments on the modal shift in Tehran, Iran

Ali Shirzadi Babakan, Abbas Alimohammadi, and Mohammad Taleai

**Abstract**

Changes in travel modes used by people, particularly reduction of the private car use, is an important determinant of effectiveness of transportation plans. Because of dependencies between the choices of residential location and travel mode, integrated modeling of these choices has been proposed by some researchers. In this paper, an agent-based microsimulation model has been developed to evaluate impacts of different transport development plans on choices of residential location and commuting mode of tenant households in Tehran, the capital of Iran. In the proposed model, households are considered as agents who select their desired residential location using a constrained NSGA-II algorithm and in a competition with other households. In addition, they choose their commuting mode by applying a multi-criteria decision making method. Afterwards, effects of development of a new highway, subway, and bus rapid transit (BRT) line on their residential location and commuting mode choices are evaluated. Results show that despite the residential self-selection effects, these plans result in considerable changes in the commuting mode of different socio-economic categories of households. Development of the new subway line shows promising results by reducing the private car use among the all socioeconomic categories of households. But the new highway development unsatisfactorily results in increase in the private car use. In addition, development of the new BRT line does not show significant effects on the commuting mode change, particularly on decrease of the private car use.

## 1. Introduction

There is a strong and complex relationship between the choices of residential location and travel mode. Necessity of simultaneous modeling of these choices in an integrated framework has been realized by many researchers (e.g. Lerman 1976; Næss 2013; Nurlaela and Curtis 2012; Vega and Reynolds-Feighan 2009). Within this framework, the residential 'self-selection' with respect to travel behavior can be considered. This type of residential self-selection is referred to the tendency of people to choose a particular neighborhood according to their travel abilities, needs, and preferences (Mokhtarian and Cao 2008; Van Wee 2009). For example, people with strong preferences for use of the public transport services usually tend to live in regions with higher accessibilities to these services. On the other hand, those who prefer to use the private car for their daily travels usually tend to live in regions with minimum restrictions on private car use. However, when a car driving lover moves to a region with a good accessibility to public transport services, he/she will be exposed to advantages of using these services and may change his/her travel behavior.

Land use and transportation plans can have considerable impacts on the built environment and subsequently on the travel behavior of people (Cao, Mokhtarian, and Handy 2007; Ewing and Cervero 2010; Guo and Chen 2007; Pinjari et al. 2007; Haddad et al. 2011; van de Walle 2009; Rand 2011). Some researchers have found that these impacts are subject to spurious effects of the residential self-selection (e.g. Cao, Mokhtarian, and Handy 2009; Handy, Cao, and Mokhtarian 2005; Mokhtarian and Cao 2008; Van Wee 2009). In other words, land use and transportation plans may not result in change of the travel behavior of people, but they may only result in movement of some people to neighborhoods of their desired transport services. As a result, ignoring the residential self-selection may lead to an overestimation of travel mode changes. For example, after development of a new subway in an area, models ignoring the residential self-selection may overestimate the contribution of subway in people's travel mode choice. Therefore, consideration of the residential self-selection is prerequisite for evaluating the impacts of land use and transportation plans on the travel behavior of people.

Majority of researches in the fields of residential location and travel mode choices have used the discrete choice models (e.g. multinomial logit (MNL), nested logit (NL), generalized extra value (GEV), mixed logit, and probit) based on the random utility maximization theory (e.g. Nurlaela and Curtis 2012; Vega and Reynolds-Feighan 2009; Rashidi, Auld, and Mohammadian 2012; Sener, Pendyala, and Bhat 2011). These conventional models are mainly based on the aggregated characteristics of people (Crooks and Heppenstall 2012; Ettema 2011). However, researchers are usually interested to investigate the impacts of land use and transportation plans on the behavior of individual households, persons or subgroups. One of the efficient and most widely used tools in this area is microsimulation. However, microsimulation models (MSMs) have some limitations in modeling of individuals' behavior in terms of their preferences, decisions, plans, and interactions. These limitations can be considerably mitigated by the integration of MSMs with individual-based models such as the agent-based models (ABMs) (Birkin and Wu 2012; Crooks and Heppenstall 2012).

ABMs are suitable for the simulation of situations where there are a large number of heterogeneous individuals with different behaviors (Davidsson 2001). ABMs are bottom-up approaches that attempt to model the complex systems at the level of individuals presented as 'agents'. Agents are the fundamental elements of these models who have different characteristics and act and interact with each other based on their perceptions of the environment (Birkin and Wu 2012; Parker et al. 2003). There is no overall agreement on definition of the term 'agent', but most agents have several common features such as the autonomy, heterogeneity, and activity (Crooks and Heppenstall 2012; Wooldridge and Jennings 1995). Behaviors of agents and their relationships with other agents and the surrounding environment are generally modeled by a set of rules. These rules are typically derived from the existing literature, expert knowledge, data surveys or empirical works (Crooks and Heppenstall 2012). Increasingly, several researchers (e.g. Benenson 1998; Devisch et al. 2009; Ettema 2011; Haase, Lautenbach, and Seppelt 2010; Waddell et al. 2003; Gaube and Remesch 2013; Salvini and Miller 2005; Veldhuisen, Timmermans, and Kapoen 2000) have used agent-based and microsimulation models for studying the residential location and travel mode choices.

The main purpose of this paper is evaluating the impacts of transport developments on the residential location and commuting mode choices of individual tenant households in Tehran, the capital of Iran. For this purpose, an agent-based microsimulation model is developed to simulate these choices. As far as the authors know, utilizing of ABMs for the evaluation of these impacts has not been considered by previous works. Therefore, the main contribution of this study is the exploration of relationships between transport development, residential self-selection, and modal shift behavior of individual households using an agent-based microsimulation model. Also, in this paper, a novel framework is proposed for choosing the residential location and commuting mode by households. In contrast with previous works, which mainly use the discrete choice models based on the random utility maximization theory, in this framework, an evolutionary multi-objective decision making algorithm (NSGA-II) combined with a competition method is used for residential location choice and a multi-criteria decision making method is used for commuting mode choice. By considering different objectives, criteria and preferences for each agent, the proposed choice mechanism leads to better consideration of the heterogeneity among agents.

Due to the lack of efficient public transport services and cheap prices of fuel in Tehran, most people prefer to use the private car in their daily travels. Therefore, different land use and transportation plans are annually prepared in Tehran with the aim of reducing the dependency of people to private car use and encouraging them to use public transport services. For this purpose, urban and transportation planners need a simulation tool to estimate the effectiveness of their plans for reducing the private car use. Although some researchers, such as Jokar Arsanjani, Helbich, and de Noronha Vaz (2013); and Meshkini and Rahimi (2011), attempt to simulate the spatial and/or temporal urban sprawl patterns in Tehran, they do not consider effects of future land use and transportation plans on the residential location and commuting mode choices. Therefore, the proposed microsimulation model in this paper is the first attempt to simulate these effects at the individual level in Tehran. This model considerably helps urban and transportation planners to simulate and predict residential self-selection and commuting mode changes resulting from the future implementation of the transport development plans (TDPs) in Tehran.

The remainder of this paper is organized as follows. An overview of the available transportation networks and commuting modes in Tehran is provided in section 2. Section 3 describes the proposed model in details. Results of the implementation and validation of the proposed model are discussed in section 4. In section 5, three transportation projects of the future development plan of Tehran are simulated and resulting changes in the residential location and commuting mode of agents are examined. Finally, major implications of the research findings are presented in the concluding section.

## 2. Transportation and commuting modes in Tehran

This study is conducted in Tehran, the capital of Iran. Tehran with an area of about 750 square kilometers and a population of about 8.3 million is located at 51° 8´ to 51° 37´ longitude and 35° 34´ to 35° 50´ latitude. This metropolis is comprised of extensive transportation networks including about 550 kilometers of highways, a subway network with 4 active intraurban lines and 72 stations with the length of about 125 kilometers, 250 bus lines with the length of about 3000 kilometers, and 6 bus rapid transit (BRT) lines with the length of about 102 kilometers (Tehran Municipality 2013). Tehran is divided into 560 traffic analysis zones (TAZs) which are used as the spatial units in this study. Figure 1 shows the existing transportation networks and TAZs in Tehran.

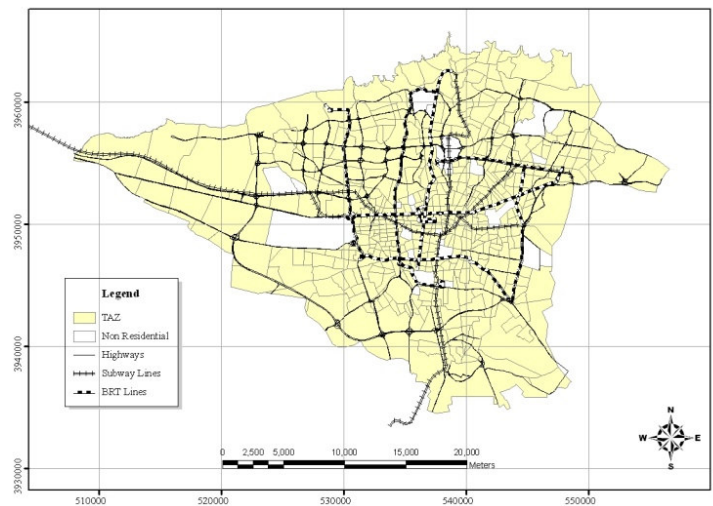
Figure 1: Existing transportation networks and TAZs in Tehran

Over the past three decades, Tehran has experienced a dramatic growth in urban population as well as the car ownership. This growth coupled with the poor urban planning and inefficient public transport services have led to significant problems such as the traffic congestion and air pollution. Although there are extensive road networks in Tehran, their capacities are not enough to accommodate this increasing number of vehicles especially during the peak hours of traffic. This paper limits its study to the work travel (commuting), because it is one of the important daily travel types, encompassing all transport modes, and results in peak-hours traffic. There are five common transport modes including the private car, subway, bus, BRT, and local taxi for commuting of people in Tehran. Shares of these transport modes in commuting of Tehran's people are shown in Figure 2. As illustrated in this Figure, other modes such as the walking, bicycle, call taxi and institutional services have minor shares in commuting of people.

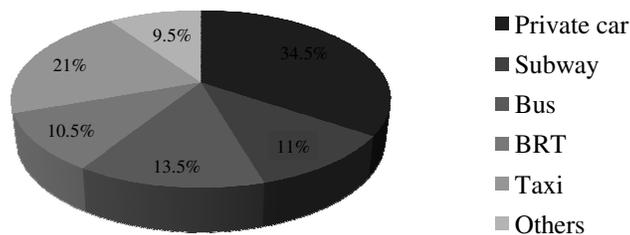

Figure 2: Shares of different transport modes in commuting of Tehran's people (Tehran Municipality 2013)

**3. Proposed model**

In this paper, an agent-based microsimulation model is developed to evaluate the impacts of different TDPs on the joint choices of residential location and commuting mode of tenant households in Tehran, the capital of Iran. In the proposed model, households are generated using the Monte Carlo simulation. They are considered as agents who select their optimal residential alternatives from the available residential zones using a constrained NSGA-II algorithm (nondominated sorting genetic algorithm II). NSGA-II is a fast and efficient multi-objective decision making algorithm in which multiple conflicting objectives are considered for residential location choice of agents according to their demographic and socioeconomic characteristics and also their criteria and preferences. This leads to consideration of more heterogeneity among agents and thereby more precise simulation of the residential location choice. Afterwards, agents compete with each other in different time periods to select a final residence from their residential alternatives. After determination of the residence of agents, their employed members are considered as the new agents who select the best available commuting mode based on their preferences to different criteria such as the travel time, travel cost, and comfortableness using a multi-criteria decision making method. Finally, different TDPs are simulated and resulting changes in the residential self-selection and commuting mode choice of agents are examined. Main components of the proposed model are shown in Figure 3.

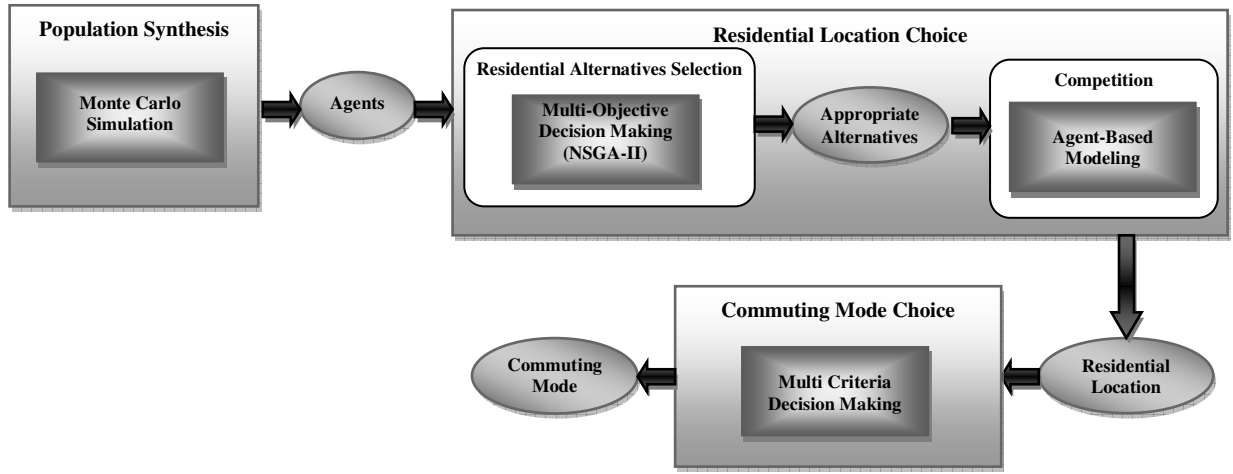

Figure 3: Main components of the proposed model

## 3.1. Population synthesis

A population of tenant households (agents) is generated using the Monte Carlo simulation such that aggregation of their attributes matches with the available demographic data. Attributes of agents are generated in a sequential approach as used by Miller et al. (2004). In this approach, initially the monthly income, number and age of the members of agents are generated. Then, the former residential zone, the number of employed members and their professional category, the number of owned cars, and the required residential area of agents are generated based on the previously determined attributes. Afterwards, the workplace of employed members of agents is randomly allocated to a zone by using the available rates of different employment categories in TAZs. Finally, criteria and preferences of agents and their employed members for choosing their residential location and commuting mode are generated according to the previously determined attributes and a survey sample consisting of stated preferences of 1580 tenant households in Tehran. This survey was conducted by filling the questionnaires in February 2012, where households with different socioeconomic characteristics stated their criteria and preferences for choosing their residential location and commuting mode in the scale range of 0 (unimportant) to 9 (very important). Table 1 presents the major criteria and the percentage of households in different socioeconomic categories who assigned a preference number greater than 4 to each criterion.

Table 1: A summary of the residential and commuting criteria and preferences of 1580 surveyed tenant households in Tehran

| Attribute | Category | Percentage | Residential location choice criteria (%) | | | | | | | | | | | | | Commuting mode choice criteria (%) | | | | | |
|---|---|---|---|---|---|---|---|---|---|---|---|---|---|---|---|---|---|---|---|---|---|
| | | | Housing Rent | Accessibility to Educational Locations | Accessibility to Commercial Locations | Accessibility to Green and Recreational Locations | Accessibility to Cultural Locations | Accessibility to Remedial Locations | Accessibility to Highways | Accessibility to Subway Stations | Accessibility to Bus Stops | Air and Noise Pollutions | Distance from the Workplace | Distance from the Former Residence | Without Traffic Restrictions | Commuting Cost | In-vehicle time | Out-of-vehicle time | Comfortability | Security | Reliability |
| Size | Single | 5.6 | 100 | 29.5 | 11.4 | 9.1 | 9.1 | 5.7 | 70.5 | 39.8 | 12.5 | 34.1 | 92.0 | 37.5 | 60.2 | 100 | 100 | 100 | 96.6 | 25.0 | 56.8 |
| | Couple | 28.1 | 100 | 8.6 | 38.3 | 41.7 | 6.3 | 22.3 | 82.7 | 48.6 | 18.0 | 46.6 | 89.9 | 50.7 | 61.3 | 99.1 | 100 | 100 | 93.0 | 27.7 | 46.8 |
| | 3-4 | 54.8 | 100 | 66.9 | 35.6 | 59.9 | 9.4 | 35.8 | 83.4 | 55.3 | 21.0 | 43.2 | 85.5 | 59.8 | 55.5 | 99.3 | 100 | 100 | 90.4 | 26.2 | 45.2 |
| | >4 | 11.5 | 100 | 92.9 | 31.3 | 63.2 | 5.5 | 44.5 | 82.4 | 59.3 | 21.4 | 43.4 | 82.4 | 59.3 | 54.9 | 100 | 100 | 100 | 90.7 | 26.9 | 49.5 |
| Monthly income (million IRR)* | < 10 | 17.6 | 100 | 45.0 | 17.6 | 36.0 | 5.0 | 20.5 | 67.6 | 57.9 | 32.7 | 23.4 | 93.9 | 65.5 | 43.2 | 100 | 100 | 100 | 80.2 | 19.1 | 37.4 |
| | 10-25 | 63.1 | 100 | 59.9 | 36.6 | 52.3 | 8.3 | 34.1 | 84.5 | 53.8 | 18.7 | 42.2 | 86.7 | 54.7 | 54.9 | 100 | 100 | 100 | 92.1 | 26.2 | 46.9 |
| | > 25 | 19.3 | 100 | 29.5 | 43.0 | 67.5 | 9.8 | 32.1 | 88.9 | 46.2 | 11.5 | 66.9 | 80.3 | 51.5 | 78.4 | 96.7 | 100 | 100 | 100.0 | 35.1 | 54.8 |
| Number of Cars | 0 | 10.8 | 100 | 66.7 | 45.0 | 51.5 | 10.5 | 45.6 | 22.8 | 83.6 | 62.6 | 30.4 | 95.3 | 64.9 | 5.3 | 100 | 100 | 100 | 78.9 | 18.1 | 45.0 |
| | 1 | 68.8 | 100 | 56.1 | 34.8 | 55.5 | 7.8 | 32.7 | 86.5 | 54.6 | 17.6 | 40.7 | 86.3 | 54.1 | 53.4 | 99.6 | 100 | 100 | 91.1 | 26.1 | 45.5 |
| | >1 | 20.4 | 100 | 27.3 | 28.0 | 42.2 | 7.5 | 19.3 | 100 | 31.4 | 4.3 | 60.9 | 83.5 | 57.5 | 98.4 | 98.1 | 100 | 100 | 99.7 | 32.9 | 51.9 |
| Total | | 100 | 100 | 51.4 | 34.5 | 52.3 | 8.0 | 31.3 | 82.3 | 53.0 | 19.7 | 43.7 | 86.7 | 55.9 | 57.3 | 99.4 | 100 | 100 | 91.5 | 26.6 | 46.8 |

\* Note: IRR (Iranian Rial), At the time of this study 1$ equivalent to 31500 IRR.

### 3.2. Residential alternatives selection

In the real world, households do not search all zones to select their residence, but they limit their search space to few desired zones according to their characteristics, requirements and preferred criteria. This fact is followed for the residential location choice of agents in the proposed model. Thus, using a constrained NSGA-II algorithm, the residential choice set of agents is limited to up to 10 optimal residential alternatives among the all TAZs. The NSGA-II algorithm, developed by Deb et al. (2002), is a fast and efficient multi-objective decision making method for the selection of the best solutions among the many available solutions. Details of this algorithm can be found in the study by Deb et al. (2002). According to the survey sample of stated preferences of tenant households in Tehran, the following rules are used as different objective functions in the NSGA-II algorithm. Depending on their criteria and preferences, agents may consider all or some of these rules in the process of their residential location choice.

*rule1:* Agents select zones as their residential alternatives in which the housing rent per square meter is compatible with their affordability and required residential area. For this purpose, maximum and minimum percentages of the monthly income which agents are willing to pay for renting a residence are simulated by the Monte Carlo simulation approach as described in section 3.1. Therefore, the search space of agents is limited to zones in which rent of their required residential area is between the specified minimum and maximum percentages of their monthly income. It should be noted that the main reason for considering the minimum limit is that many agents, particularly those with high incomes, usually prefer to reside in neighborhoods with qualified cultural and social characteristics.

*rule2:* Agents prefer to select zones in closer distances to the workplaces of their employed members.

*rule3:* Agents prefer to select zones in closer distance from their former residential area.

*rule4:* Agents prefer to select zones with minimum environmental pollutions. A constraint is considered for agents with very important preferences for air and noise pollutions, as they only search among zones in which pollution levels are medium to clean.

***rule5:*** Agents prefer to select zones with no traffic restrictions on the private car use. A constraint is considered for agents with very important preferences for traffic restrictions, as they only search among the non-restricted zones.

***rule6:*** Agents prefer to select zones with higher accessibilities to public facilities including the commercial, educational, green, recreational, remedial, and cultural facilities. Accessibility of zones to the public facilities for each agent is calculated by Eq. (1) (Tsou, Hung, and Chang 2005).

$$A_i^a = \sum_f \sum_{j_f} p_f^a * w_{j_f} * d_{ij_f}^{-2} \qquad \text{Eq. (1)}$$

where:

$A_i^a$ is the accessibility of zone *i* to the public facilities for agent *a*;

*f* is the type of public facility;

$j_f$ is the *j*th case of the *f*th type of public facility;

$p_f^a$ is the preference of agent *a* to the type of public facility *f* which is simulated by the Monte Carlo simulation method, where $\sum p_f^a = 1$;

$w_{j_f}$ is the relative effect of $j_f$ which is calculated by $w_{j_f} = A_{j_f}/\max(A_{j_f})$, where $A_{j_f}$ is the area of $j_f$;

$d_{ij_f}$ is the distance between zone *i* and the public facility $j_f$.

***rule7:*** Agents prefer to select zones with higher accessibilities to transportation services including the highways and/or public transport services (bus and subway stations). Accessibility of zones to the transportation services for each agent is calculated by Eq. (2) (Currie 2010; Delbosc and Currie 2011).

$$At_i^a = \sum_t \sum_{j_t} p_t^a * \frac{A_{j_t}}{A_i} \qquad \text{Eq. (2)}$$

where:

$At_i^a$ is the accessibility of zone *i* to transportation services for agent *a*;

*t* is the type of transportation service;

$j_t$ is the *j*th case of the *t*th type of transportation service;

$p_t^a$ is the preference of agent *a* to the type of transportation service *t* which is simulated by the Monte Carlo simulation method, where $\sum p_t^a = 1$;

$A_i$ is the area of zone *i*;

$A_{j_t}$ is the service area of $j_t$ which is inside the zone *i*; the service ranges of highways, subway stations, BRT, and bus stops are set to 2, 1.9, 1.7 and 1.2 km, respectively. These distances, obtained from the comprehensive transportation and traffic studies of Tehran, are slightly higher than those used in other cities, because besides walking, Tehran's people usually use cheap local taxis for access to the transport services.

### 3.3. Competition

Agents compete with each other in different time periods to choose a residence from their residential alternatives. For this purpose, they are randomly distributed in monthly time periods according to the volume of relocation in different months of the year in Tehran. In addition, the residential capacity of each zone is estimated by multiplying the ratio of residential area in that zone to all zones, the number of agents, and an equilibrium parameter in each period. The equilibrium parameter is derived from the volume of demand and supply for renting residences in different months of the year in Tehran.

Agents search their residential alternatives according to the distance from their former residence and reside in the first alternative which has the residential capacity. If the residential capacity of a zone is not enough to satisfy demands of all agents, they compete with each other for residing in that zone. In this competition, agents with fewer members (except singles), with higher incomes, and without a child would succeed, respectively. If demandants are equal with respect to the mentioned criteria, winners are randomly selected and defeated agents try to reside in their next alternatives. This process is continued until all agents reside in one of their residential alternatives or all alternatives of the defeated agents are evaluated. If a number of agents cannot reside in any of their alternatives, they are moved to the competition process in the next time period. If they still cannot reside in any zone, no residence is allocated for them. Generally, they are agents with low incomes who cannot afford to rent a residence in any zone and they probably have to move to suburbs.

### 3.4. Commuting mode choice

After determination of the residential location of agents, the employed members of agents are considered as the new agents who select the best available commuting mode based on their criteria and preferences using a multi-criteria decision making method. The most important criteria for the commuting mode choice are derived from the survey sample of stated preferences of commuting mode choice in Tehran. These criteria include the travel cost, out-of-vehicle time, in-vehicle time, comfortability, security, and reliability. It is requested from a number of transportation experts to evaluate the transport modes including the walking, private car, bus, BRT, subway, and taxi with respect to these criteria at the daily peak hours of traffic in Tehran. They relatively compare the transport modes with respect to each criterion and assign a score in the scale range of 1 to 10. The average of normalized scores of experts is used as the score of each transport mode with respect to each criterion. Agents have their own preferences with respect to these criteria. These preferences are normalized and used for the calculation of the suitability of each transport mode for each agent as follows.

$$S_m^a = \sum_{c=1}^{n} p_c^a * v_m^c \qquad \text{Eq. (3)}$$

where:

$S_m^a$ is the suitability of the transport mode *m* for agent *a*;

$p_c^a$ is the preference of agent *a* to the criterion *c*, where $\sum p_c^a = 1$;

$v_m^c$ is the score of transport mode *m* with respect to the criterion *c*.

Finally, agents choose the transport mode with the maximum suitability for their commuting. It should be noted that the level-of-service (LOS) attributes including the travel cost, in-vehicle and out-of-vehicle times for different transport modes are obtained from the shortest path between the residence and workplace of agents on the network models in GIS.

## 4. Implementation and validation

The proposed model was implemented by the MATLAB software version 7.12. Using the Monte Carlo simulation, 50000 tenant households with different socio-economic characteristics were generated as agents. They searched among TAZs and selected their optimal residential alternatives using the NSGA-II algorithm. Then, they chose a final residence from their alternatives by the aforementioned competition mechanism. Figure 4 shows the distribution of residential locations of agents in TAZs. Finally, 58941 employed members of the simulated households selected their commuting mode according to Equation (3). Selected modes by different socioeconomic categories of agents are presented in Table 2.

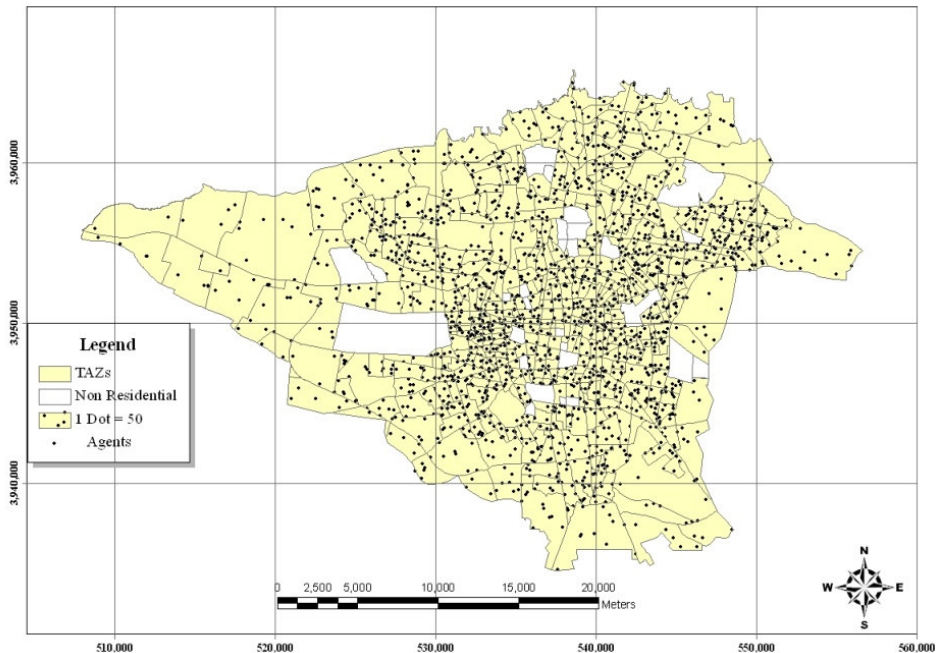

Figure 4: Distribution of residential zones of the simulated agents in Tehran

Table 2: Commuting modes of different socioeconomic categories of the simulated agents in Tehran

| Attribute | Category | Number | Percentage | Commuting mode (%) | | | | | |
|---|---|---|---|---|---|---|---|---|---|
| | | | | Private Car | Subway | Bus | BRT | Taxi | Walking |
| Gender | Female | 18104 | 30.7 | 39.1 | 12.9 | 13.7 | 10.6 | 22.5 | 1.2 |
| | Male | 40837 | 69.3 | 38.5 | 12.2 | 14.1 | 10.9 | 22.4 | 1.9 |
| Household Size | Single | 2888 | 4.9 | 47.5 | 10.1 | 9.8 | 7.8 | 23.7 | 1.1 |
| | Couple | 17394 | 29.5 | 40.1 | 11.8 | 13.0 | 10.0 | 23.6 | 1.5 |
| | 3-4 | 31636 | 53.7 | 37.9 | 12.7 | 14.5 | 11.3 | 21.8 | 1.9 |
| | > 4 | 7023 | 11.9 | 35.1 | 13.6 | 15.9 | 11.8 | 22.0 | 1.6 |
| Average Monthly Income of Household (million IRR*) | < 10 | 10597 | 18.0 | 29.2 | 17.8 | 19.1 | 15.7 | 16.8 | 1.4 |
| | 10-25 | 37331 | 63.3 | 37.7 | 11.9 | 14.7 | 10.8 | 23.1 | 1.8 |
| | > 25 | 11013 | 18.7 | 51.2 | 9.0 | 6.5 | 6.1 | 25.6 | 1.6 |
| Number of Private Cars of Household | 0 | 6532 | 11.1 | 0.0 | 25.7 | 23.4 | 20.6 | 28.5 | 1.8 |
| | 1 | 40050 | 67.9 | 34.8 | 12.2 | 15.7 | 11.7 | 23.8 | 1.8 |
| | > 1 | 12359 | 21.0 | 71.6 | 6.1 | 3.5 | 2.9 | 14.7 | 1.2 |
| Commuting Distance (km) | < 5 | 9607 | 16.3 | 21.7 | 11.7 | 15.7 | 12.0 | 28.6 | 10.3 |
| | 5 - 15 | 37782 | 64.1 | 36.7 | 10.8 | 17.1 | 12.7 | 22.6 | 0.0 |
| | > 15 | 11552 | 19.6 | 59.4 | 18.2 | 2.3 | 3.5 | 16.6 | 0.0 |
| Total | | 58941 | 100 | 38.7 | 12.4 | 14.0 | 10.8 | 22.4 | 1.7 |

One of the greatest challenges of utilizing the ABMs is their validation (Crooks and Heppenstall 2012). In this research, a survey sample of 1485 actual tenant households was used for validation of the proposed model. These data were not used in the population synthesis of agents. Residence and commuting mode of these households were simulated by the proposed model and results were compared with their actual residence and commuting mode. Validation results for different socioeconomic categories of sample households are presented in Table 3. As shown in this table, the proposed model correctly simulates the residential zone of 62.6% of households and the commuting mode of 69.6% of their employed members. Also, the model properly simulates the commuting mode of 78.0% of females and 76.8% of all households who use the private car for their commuting. Therefore, the proposed model seems to show a good and promising performance.

Table 3: Validation results of the proposed model

| Attribute | Category | Number | Percentage | Correctly Simulated Residential Zone (%) | * Correctly Simulated Residential Neighborhood (%) | Correctly Simulated Commuting Mode (%) | | | | | |
|---|---|---|---|---|---|---|---|---|---|---|---|
| | | | | | | Private Car | Subway | Bus | BRT | Taxi | Walking |
| Gender | Female | 391 | 26.3 | 59.8 | 78.8 | 78.0 | 67.3 | 61.7 | 61.1 | 65.6 | 75.0 |
| | Male | 1094 | 73.7 | 63.5 | 82.4 | 76.3 | 68.3 | 62.9 | 63.5 | 66.0 | 66.7 |
| Household Size | Single | 58 | 3.9 | 70.7 | 87.9 | 79.3 | 60.0 | 50.0 | 50.0 | 66.7 | 0.0 |
| | Couple | 475 | 32.0 | 64.2 | 83.2 | 77.4 | 65.0 | 61.5 | 60.4 | 70.9 | 62.5 |
| | 3-4 | 820 | 55.2 | 62.2 | 80.5 | 76.6 | 69.8 | 64.2 | 65.2 | 63.8 | 75.0 |
| | > 4 | 132 | 8.9 | 55.3 | 78.8 | 73.5 | 70.0 | 58.3 | 62.5 | 62.5 | 66.7 |
| Average Monthly Income of Household (million IRR) | < 10 | 247 | 16.6 | 64.4 | 86.2 | 70.8 | 68.0 | 62.7 | 63.9 | 63.4 | 50.0 |
| | 10-25 | 937 | 63.1 | 60.3 | 80.7 | 75.3 | 68.6 | 62.8 | 62.6 | 67.2 | 70.0 |
| | > 25 | 301 | 20.3 | 68.1 | 80.1 | 82.0 | 65.2 | 60.0 | 63.2 | 63.0 | 75.0 |
| Number of Private Cars of Household | 0 | 147 | 9.9 | 62.6 | 78.9 | 100 | 76.7 | 70.0 | 62.5 | 67.6 | 60.0 |
| | 1 | 1059 | 71.3 | 62.7 | 82.7 | 72.0 | 66.7 | 61.8 | 63.2 | 65.9 | 68.2 |
| | > 1 | 279 | 18.8 | 62.0 | 78.1 | 85.0 | 56.3 | 40.0 | 60.0 | 64.4 | 100 |
| Commuting Distance (km) | < 5 | 205 | 13.8 | 64.9 | 86.8 | 57.6 | 72.2 | 71.4 | 74.1 | 78.1 | 67.9 |
| | 5 - 15 | 968 | 65.2 | 62.5 | 81.5 | 74.4 | 55.2 | 61.4 | 61.4 | 65.6 | 100 |
| | > 15 | 312 | 21.0 | 61.2 | 77.9 | 84.3 | 86.8 | 42.9 | 50.0 | 47.4 | 100 |
| Total | | 1485 | 100 | 62.6 | 81.5 | 76.8 | 68.1 | 62.6 | 63.0 | 65.9 | 67.9 |

* Note: A neighborhood with Euclidean distance of 5 kilometers

## 5. Impact assessment of the TDPs

In this section, impacts of three TDPs including construction of a new highway, BRT, and subway on the residential self-selection and commuting mode choice of tenant households are evaluated. These TDPs are the major under construction components of the future development plan of Tehran and will be operational in the near future (Figure 5). In this research, these TDPs are separately simulated and their effects on the accessibility and housing rent are estimated. Then, the simulated agents reselect their residence and commuting mode under the new conditions. Since all other conditions are assumed to be constant and only the transportation system is changed, the observed changes in the residential location and commuting mode of agents can be attributed to the impacts of TDPs.

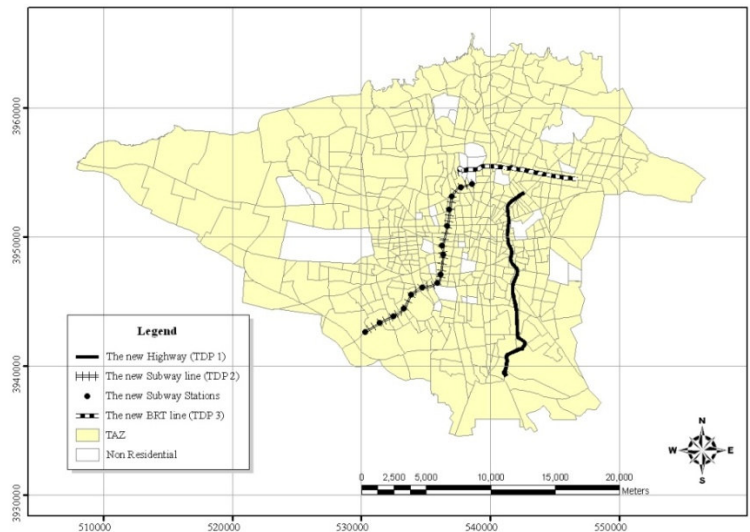

Figure 5: Location of the new transport development plans (TDPs) in Tehran

TDPs influence many components of a city including the accessibility, travel behavior, traffic, environmental pollutions, land uses, residential location, and housing price in the long- and/or short-time periods. However, in this research, only the impacts of TDPs on the accessibility, travel behavior, housing rent, and residential location choice are investigated. In other words, it is assumed that these TDPs do not have any significant impacts on other determinants of the residential location and commuting mode choices.

TDPs significantly improve the accessibility of their neighborhoods to the transport services. The accessibility of TAZs to the transport services is updated using Eq. (2). As a result, due to improvement of the accessibility, the housing rent is also changed in these neighborhoods. Consultations with real-estate agencies besides the experiences obtained from the construction of previous TDPs in Tehran show that the highway and subway developments can have significant impacts on the housing rent, but the BRT developments have not considerable effects on the housing rent. Therefore, the housing rents within the neighborhoods of the highway and subway developments are updated using the following equation:

$$R_i^n = \frac{\left(A_i - \sum A_i^{rr'}\right)R_i + \sum(A_i^{rr'} G_{rr'} R_i)}{A_i} \qquad \text{Eq. (4)}$$

where:

$R_i^n$ and $R_i$ are the new and previous housing rents per square meter in zone $i$, respectively;

$r$ and $r'$ are the beginning and ending radiuses of a ring buffer around the new highway or subway;

$A_i$ and $A_i^{rr'}$ are the areas of zone $i$ and the ring buffer with radiuses $r$ and $r'$, respectively;

$G_{rr'}$ is the change rate of housing rent in the buffer area. The values of this parameter are empirically estimated by consultation with the real-estate agencies and by experiences obtained from the construction of previous TDPs in Tehran.

For studying how the residential location and commuting mode of agents are changed by TDPs, the proposed model was repeatedly run and changes of the residential location and commuting mode of agents were observed. A summary of these changes is presented in Table 4. Generally, considerable changes in the commuting mode of agents are observed in neighborhoods of TDPs. However, a part of these changes is due to residential self-selection of some agents with respect to their commuting preferences. In other words, some agents with high interests to commute by the private car, subway or BRT move to vicinities of the new highway, subway or BRT, respectively. Therefore, increase in the ridership of subway or BRT in neighborhoods of the new developments does not necessarily indicate the effectiveness of these plans. But what is important is that which and how many agents change their commuting mode to the developed modes. Results of these investigations are shown in Table 4.

Interesting results are obtained from simulation of the residential location and commuting mode choices of agents after implementation of the new highway (TDP 1). Private car use significantly increases in the neighboring zones of the new highway (Figure 6). This increase is greater in the northern neighboring zones and indicates that residents of these zones are more interested to use the private car, because the mean car ownership in these zones is more than those of the other neighboring zones. However, private car use also shows a decreasing trend in some zones, because a number of agents with high interests to private car use move from these zones to the neighborhoods of the new highway (residential self-selection effects). Although increase in the private car use in the neighboring zones of the new highway is partly due to the residential self-selection, some agents also change their commuting mode to the private car. As presented in Table 4, development of the new highway generally results in 1.1% increase in the private car use. On the other hand, the greatest decreases of the private car use are observed in the shares of subway, bus and BRT, respectively. Figure 7 shows increase in the private car use by different socioeconomic categories of agents. As can be seen, after implementation of the new highway, agents with fewer members, higher incomes, more owned cars, and longer commuting distances show more interests to change their commuting mode to the private car. As a result, development of the new highway encourages people to use their private cars and may result in increase in the traffic congestion and environmental pollutions in the long term. This completely conflicts with the main goal of urban and transportation planners in Tehran which is decreasing the private car use.

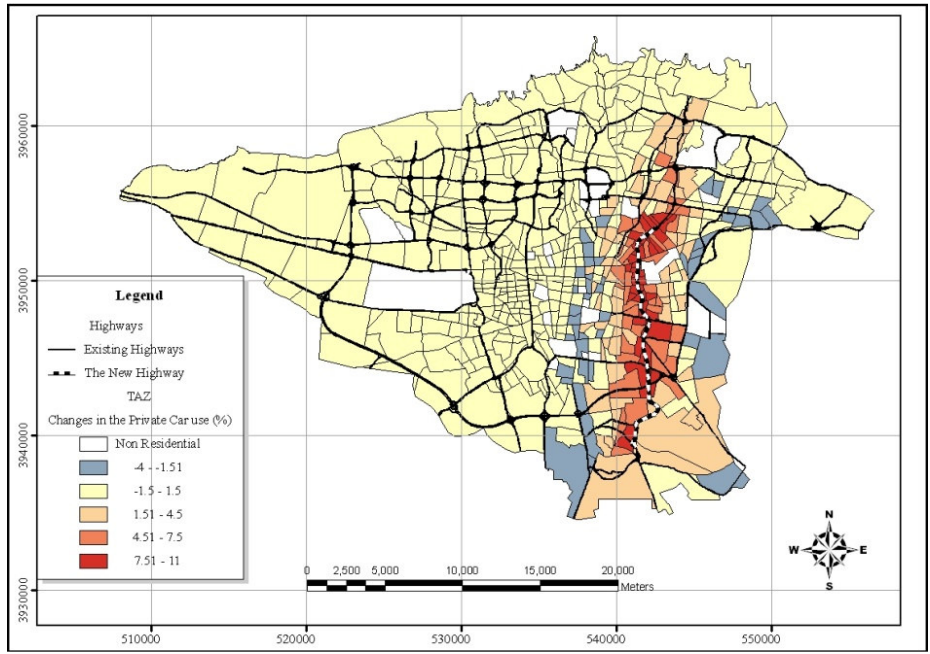

Figure 6: Changes of the private car use in TAZs after implementation of the new highway

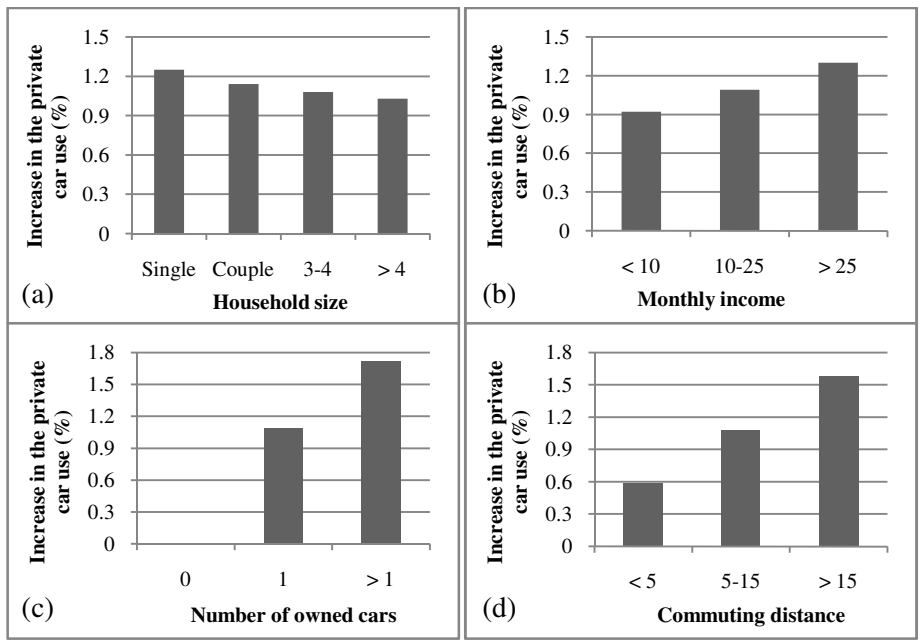

Figure 7: Increase in the private car use by agents with different household sizes (a); monthly incomes (b); number of owned cars (c); and commuting distances (d) after implementation of the new highway

By development of the new subway line (TDP 2), accessibility of many deprived areas of the southwest of Tehran to the subway network is significantly improved. This plan results in substantial changes in the commuting mode of residents in the southwest zones of Tehran. As illustrated in Figure 8, share of the subway in commuting of people increases up to 17.5% in some southwest zones. Also, due to interconnection of the subway lines, use of the subway shows slight increases in the neighborhoods of the previous stations. However, because of the residential self-selection effects, use of the subway decreases in some zones which are mainly located in north of the new line and have poor accessibilities to the subway network. In other words, a number of agents with high interests to use the subway mode move from these zones to neighborhoods of the new subway stations. In addition to residential self-selection which leads to increase in subway use in the neighboring zones of the new subway line, some agents change their commuting mode to the subway. As shown in Table 4, implementation of the new subway line results in 2.8% increase in the share of subway and 1.07% decrease in the private car use. Figure 9 shows increase in the subway use by different socioeconomic categories of agents. Although after implementation of the new subway line, all categories of agents show considerable willingness to change their commuting mode to the subway, agents with more members, lower incomes, fewer owned cars, and longer commuting distances show higher interests to change their commuting mode to the subway. Results show that development of the new subway line is significantly effective in decrease in the private car use even among the agents with high incomes and more than one owned car.

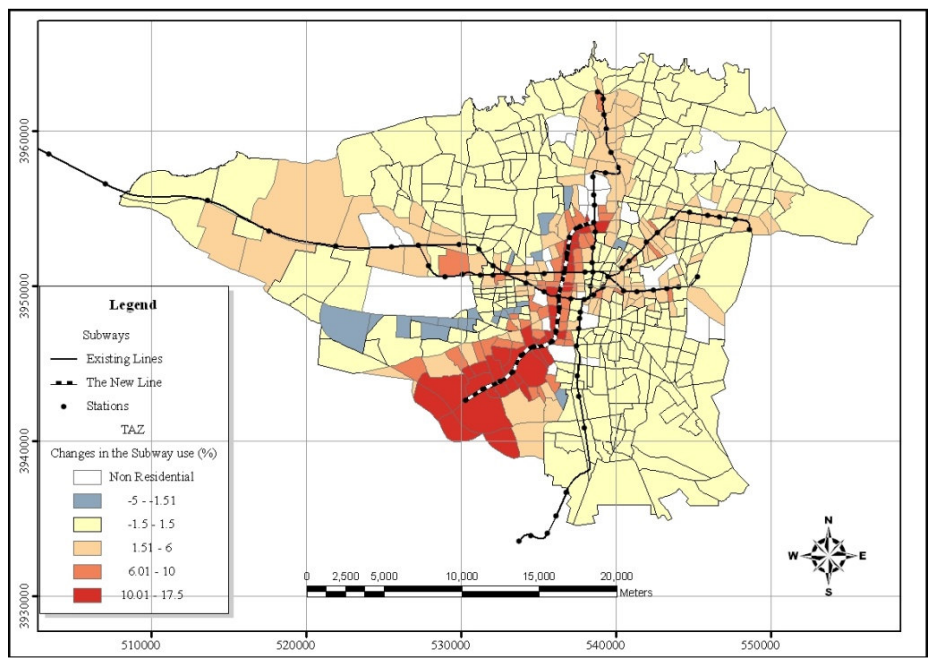

Figure 8: Changes of the subway use in TAZs after implementation of the new subway line

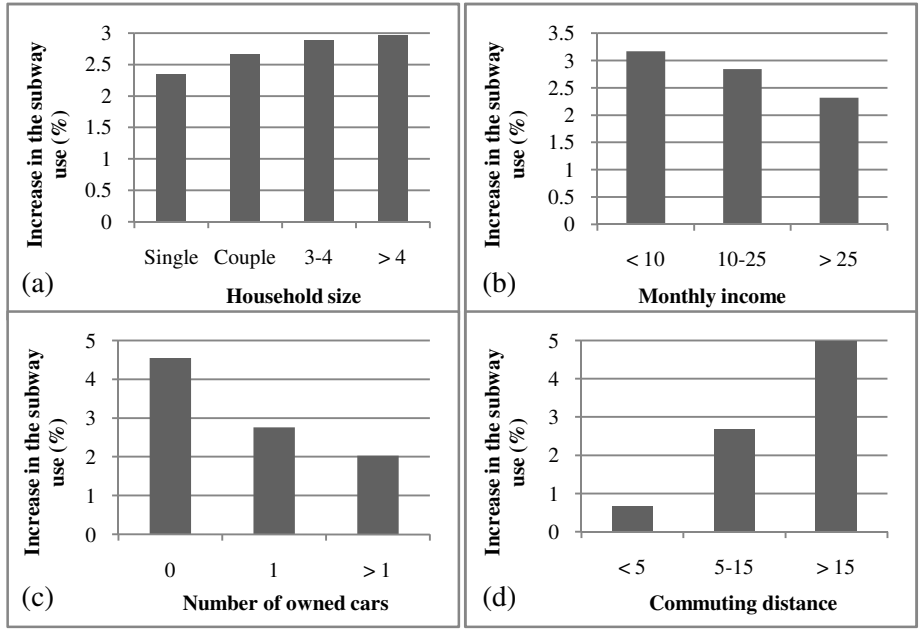

Figure 9: Increase in the subway use by agents with different household sizes (a); monthly incomes (b); number of owned cars (c); and commuting distances (d) after implementation of the new subway line

Figure 10 shows changes of the BRT use in TAZs after implementation of the new BRT line (TDP 3). These changes are greater in the southern neighboring zones of the new line. This shows that residents of these zones have more willingness to use BRT for their commuting. The main reason for this may be that the mean income and car ownership in these zones are less than those of the other neighboring zones and generally agents with low incomes and fewer owned cars show higher interests to use BRT. In contrast to increase in BRT use in the neighboring zones of the new line, its use decreases in a significant number of zones. In other words, some agents with high interests to use BRT for their commuting move from these zones to neighborhoods of the new line. Therefore, share of the BRT decreases in these zones and increases in neighborhoods of the new line. This indicates that a major part of impacts of the new BRT line on the commuting mode changes is due to residential self-selection. As a result, total share of the BRT mode does not show a significant increase. As presented in Table 4, after implementation of the new BRT line, only 0.36% of agents change their commuting mode to the BRT. Also, share of the private car does not show a meaningful decrease. As illustrated in Figure 11, agents with more members, lower incomes, fewer owned cars, and shorter commuting distances show higher interests to change their commuting mode to BRT. However, agents with high incomes and more than one owned car do not show considerable interests to change their commuting mode to BRT. As a result, the new BRT line does not show significant efficiency to encourage agents especially those with high incomes and a high number of owned cars to change their commuting mode form the private car to the BRT mode.

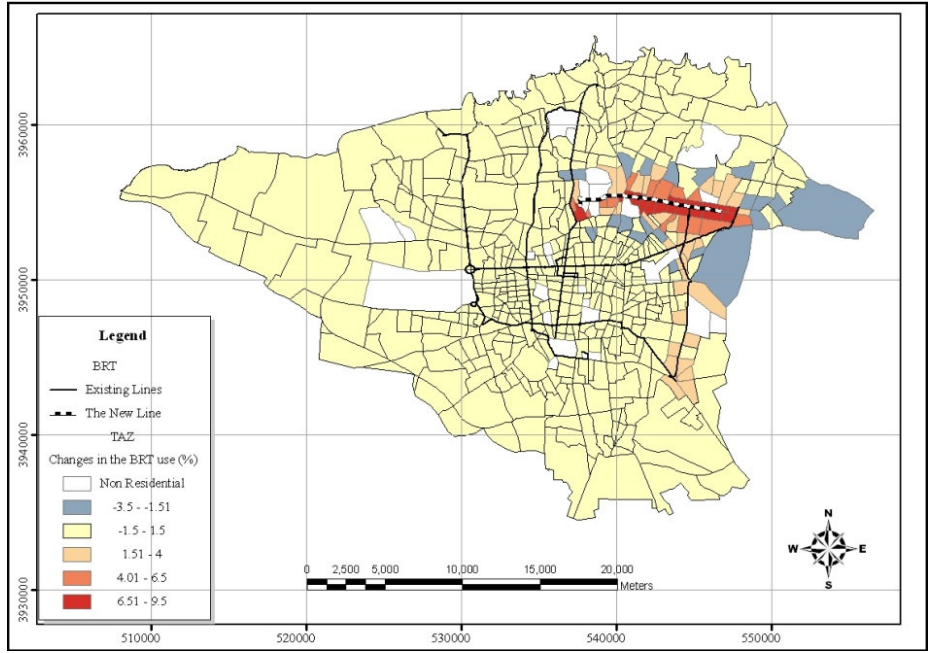

Figure 10: Changes of the BRT use in TAZs after implementation of the new BRT line

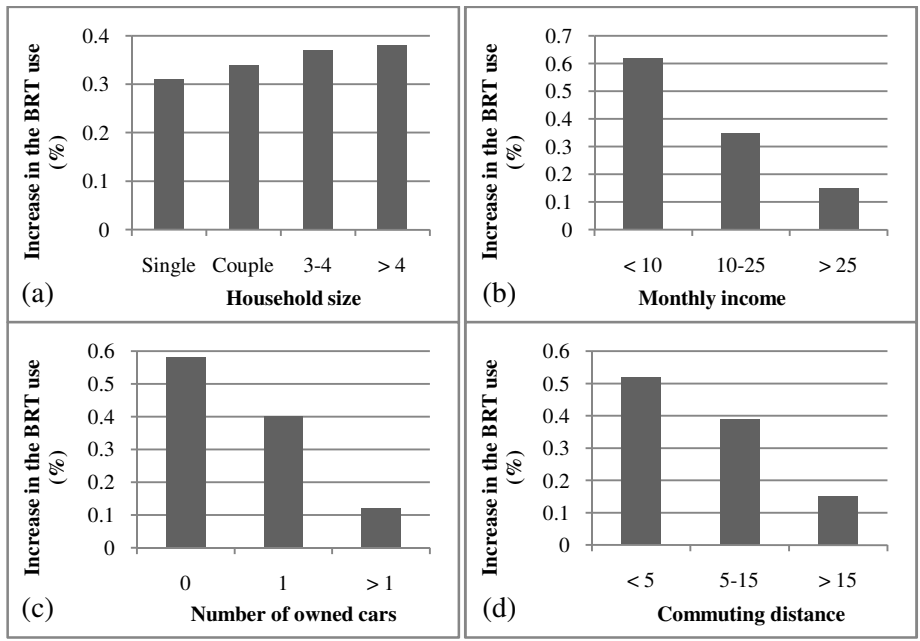

Figure 11: Increase in the BRT use by agents with different household sizes (a); monthly incomes (b); number of owned cars (c); and commuting distances (d) after implementation of the new BRT line

Table 4: Changes in the residential location and commuting mode of agents after implementation of the three transport development plans in Tehran

| Attribute | Category | Number | Percentage | Development of the new highway (TDP 1) | | | | | | | Development of the new subway line (TDP 2) | | | | | | | Development of the new BRT line (TDP 3) | | | | | | |
|---|---|---|---|---|---|---|---|---|---|---|---|---|---|---|---|---|---|---|---|---|---|---|---|---|
| | | | | Residential Zone | Private Car | Subway | Bus | BRT | Taxi | Walking | Residential Zone | Private Car | Subway | Bus | BRT | Taxi | Walking | Residential Zone | Private Car | Subway | Bus | BRT | Taxi | Walking |
| Gender | Female | 18104 | 30.7 | 3.92 | +1.08 | -0.39 | -0.30 | -0.28 | -0.10 | -0.01 | 4.08 | -1.01 | +2.73 | -0.49 | -0.56 | -0.67 | 0.00 | 0.65 | -0.07 | -0.08 | -0.04 | +0.37 | -0.17 | -0.01 |
| | Male | 40837 | 69.3 | 3.93 | +1.11 | -0.41 | -0.31 | -0.29 | -0.09 | -0.01 | 4.08 | -1.10 | +2.84 | -0.45 | -0.58 | -0.71 | 0.00 | 0.69 | -0.08 | -0.07 | -0.03 | +0.36 | -0.17 | -0.01 |
| Household Size | Single | 2888 | 4.9 | 4.47 | +1.25 | -0.48 | -0.35 | -0.31 | -0.10 | 0.00 | 4.40 | -0.76 | +2.35 | -0.42 | -0.45 | -0.73 | 0.00 | 0.76 | -0.03 | -0.07 | -0.03 | +0.31 | -0.17 | 0.00 |
| | Couple | 17394 | 29.5 | 4.15 | +1.14 | -0.42 | -0.33 | -0.28 | -0.10 | -0.01 | 4.04 | -0.98 | +2.66 | -0.42 | -0.53 | -0.74 | 0.00 | 0.64 | -0.08 | -0.07 | -0.02 | +0.34 | -0.16 | -0.01 |
| | 3-4 | 31636 | 53.7 | 3.80 | +1.08 | -0.40 | -0.28 | -0.30 | -0.09 | -0.01 | 4.11 | -1.11 | +2.89 | -0.48 | -0.63 | -0.67 | 0.00 | 0.69 | -0.07 | -0.08 | -0.04 | +0.37 | -0.18 | -0.01 |
| | > 4 | 7023 | 11.9 | 3.73 | +1.03 | -0.34 | -0.33 | -0.24 | -0.10 | -0.01 | 3.92 | -1.27 | +2.96 | -0.53 | -0.50 | -0.67 | 0.00 | 0.68 | -0.10 | -0.07 | -0.04 | +0.38 | -0.16 | -0.01 |
| Average Monthly Income of Household (million IRR) | < 10 | 10597 | 18.0 | 3.58 | +0.92 | -0.33 | -0.25 | -0.20 | -0.14 | -0.01 | 4.63 | -1.43 | +3.17 | -0.43 | -0.45 | -0.85 | 0.00 | 0.82 | -0.14 | -0.13 | -0.08 | +0.62 | -0.25 | -0.01 |
| | 10-25 | 37331 | 63.3 | 4.28 | +1.09 | -0.44 | -0.28 | -0.28 | -0.07 | -0.01 | 4.23 | -1.01 | +2.84 | -0.50 | -0.66 | -0.67 | 0.00 | 0.79 | -0.07 | -0.08 | -0.02 | +0.35 | -0.17 | -0.01 |
| | > 25 | 11013 | 18.7 | 3.10 | +1.30 | -0.35 | -0.44 | -0.37 | -0.13 | -0.01 | 3.04 | -0.93 | +2.32 | -0.37 | -0.39 | -0.64 | 0.00 | 0.16 | -0.02 | -0.03 | -0.03 | +0.15 | -0.06 | -0.02 |
| Number of Private Cars of Household | 0 | 6532 | 11.1 | 1.97 | 0.00 | -0.06 | 0.00 | 0.00 | +0.06 | 0.00 | 6.87 | 0.00 | +4.55 | -1.26 | -1.35 | -1.94 | 0.00 | 1.91 | 0.00 | -0.18 | -0.09 | +0.58 | -0.29 | -0.02 |
| | 1 | 40050 | 67.9 | 4.64 | +1.09 | -0.37 | -0.30 | -0.28 | -0.12 | -0.01 | 4.38 | -1.14 | +2.76 | -0.46 | -0.61 | -0.54 | 0.00 | 0.61 | -0.10 | -0.08 | -0.02 | +0.40 | -0.19 | -0.01 |
| | > 1 | 12359 | 21.0 | 2.65 | +1.72 | -0.71 | -0.49 | -0.45 | -0.08 | 0.00 | 1.63 | -1.41 | +2.03 | -0.06 | -0.05 | -0.52 | 0.00 | 0.27 | -0.02 | -0.02 | -0.03 | +0.12 | -0.04 | -0.01 |
| Commuting Distance (km) | < 5 | 9607 | 16.3 | 3.45 | +0.59 | -0.07 | -0.20 | -0.17 | -0.09 | -0.06 | 2.07 | -0.09 | +0.66 | -0.08 | -0.07 | -0.40 | -0.01 | 0.44 | -0.10 | -0.08 | -0.04 | +0.52 | -0.23 | -0.06 |
| | 5 - 15 | 37782 | 64.1 | 4.00 | +1.08 | -0.29 | -0.40 | -0.36 | -0.03 | 0.00 | 4.28 | -0.65 | +2.68 | -0.69 | -0.85 | -0.49 | 0.00 | 0.84 | -0.08 | -0.09 | -0.03 | +0.39 | -0.18 | 0.00 |
| | > 15 | 11552 | 19.6 | 4.11 | +1.58 | -1.06 | -0.09 | -0.14 | -0.30 | 0.00 | 5.10 | -3.28 | +4.99 | -0.04 | -0.07 | -1.60 | 0.00 | 0.36 | -0.02 | -0.02 | -0.03 | +0.15 | -0.08 | 0.00 |
| Total | | 58941 | 100 | 3.93 | +1.10 | -0.41 | -0.31 | -0.29 | -0.10 | -0.01 | 4.08 | -1.07 | +2.80 | -0.46 | -0.57 | -0.69 | 0.00 | 0.68 | -0.07 | -0.08 | -0.03 | +0.36 | -0.17 | -0.01 |

## 6. Conclusion

In this paper, the residential location choice of tenant households and the commuting mode choice of their employed members were modeled using an agent-based microsimulation model. Then, impacts of the three TDPs including development of the new highway, subway and BRT line on these choices were evaluated by repeatedly running the model. Although microsimulation models have some limitations such as data requirements and computational complexities, recent advances in computational capacities and integration with new technologies such as ABMs have reduced these limitations. Agent-based microsimulation models enable us to study the residential location choice and travel behavior of individuals. Also, it provides a robust and flexible framework for implementation of different rules, assumptions, interactions, and levels of description and aggregation. This framework allows us to better study the behavioral and causal relationships in the urban system.

Results of the present study indicate that TDPs lead to considerable effects on the commuting mode choice of households. A major part of these effects is due to the residential self-selection of households with respect to their travel preferences. After implementation of each TDP, some households who prefer to commute by the developed mode attempt to move to its neighborhood. Therefore, use of the developed mode is increased in neighborhoods of the development. As a result, it should be noted that the increase in use of the developed mode in its neighboring areas does not necessarily indicate its efficiency, because a significant part of this increase is due to the residential self-selection. In addition to residential self-selection effects, these plans lead to considerable changes in commuting mode of households. This is very important for evaluating the efficiency of different transportation plans. Results show that the new highway leads to increase in the private car use in commuting of households. In other words, after implementation of the new highway, private car use becomes more attractive for some households who previously used the public transport services. This is exactly the opposite of what urban and transportation planners desire and may result in more traffic, air and noise pollutions in the long term. But, results of the new subway line development are very promising, because it leads to decrease in the private car use. Some households from all socioeconomic categories, even those with high incomes and a high number of owned cars, change their commuting mode from the private car to the subway. Furthermore, development of the new BRT line results in change of the commuting mode of some households. However, these changes are more observable among households with low incomes and without the private car. In other words, households with high incomes and a high number of owned cars do not show interest to change their commuting mode from the private car to BRT. Therefore, it seems that the new subway line is more popular than the new BRT line, and it is more efficient to decrease the private cars use. It should be emphasized that results of this research are exposed to assumptions, rules and conditions of the simulation and can be regarded as the scenario-based results.

Various aspects of the proposed model can be improved in the future. First, the rules and conditions used in this study to model the behavior of agents and their relationships with other agents and the surrounding environment can be enhanced to improve the results. Second, use of the TAZs as the spatial units leads to modifiable areal unit problem (MAUP) and errors in measurement of the network-based attributes of the transport services. This means that changes in size or configuration of zones can affect the results. In order to reduce these effects and also to enhance the efficiency of the proposed model in the large-scale planning, smaller spatial units such as the census blocks, cells, and parcels can be used. In addition, effects of varying sizes and configurations of spatial units on results can be examined. Third, the proposed model can be extended to evaluate impacts of TDPs on the residential location and commuting mode choices of not only tenant households but also all households. Fourth, the proposed model can be used to evaluate impacts of different transport policies (e.g. changes in the fuel price, changes in the parking space and cost, and changes in the existing transportation networks) on residential location and commuting mode choices of households. Fifth, although commuting mode choice has been modeled in a typical day of the year in this paper, the model can be developed to consider time-related criteria including season, weather conditions, and school holidays in selection of an appropriate commuting mode. Finally, a price bidding framework and a housing supply module can be added to the proposed model in the future.